\documentclass[aps,prl,reprint,superscriptaddress,showpacs]{revtex4-2}
\usepackage{amsmath}
\usepackage{amssymb}
\usepackage{graphicx}
\usepackage{hyperref}

\begin{document}

\title{About Asymptotic continuation of Einstein's equation
 }

\author{Aritra Sanyal}
\email{aritrasanyal1@gmail.com}
\affiliation{Institute of Astronomy, Space and Earth Science, P 177, CIT Road, Kolkata 700054, West Bengal, India\\aritra@iases.org.in}

\author{Valery B. Morozov}
\email{valery_morozov@hotmail.com}
\affiliation{St. Gamalei 11-2-24, 123098, Moscow, Russia}

\date{\today}

\begin{abstract}
The energy density of the gravitational field is a full-fledged source of the gravitational field.
This principle of Einstein was not implemented by him in the Einstein equation. Not long ago,
it was possible to find the energy-momentum tensor, asymptotically equal to a part of the
Ricci tensor. This tensor was included in the Einstein equation. As a result, the Einstein
equation was reduced. Initially, the reduced gravitational field equation was solved only for
the field \(g_{00}\) of a point source. Here we obtain the full metric of the gravitational field of a	
point source using the reduced field equation. The solution is interpreted as a potential well of
extreme depth with a radius approximately one hundred and forty times greater than the
Schwarzschild radius.

\end{abstract}

\maketitle

\section{Introduction}
 Einstein introduced Riemannian space into physics as a physical object. This made it possible to describe gravity and its accompanying phenomena, at least in the region of not too large fields.The presented material does not go beyond the principles of the general theory of relativity; on the contrary, the theory managed to include a full-fledged principle of conservation of energy, which is absent in Einstein's theory of gravity.
The first results were obtained in $[1,2,3]$. In this work, the energy-momentum tensor of the gravitational field and the gravitational field equation are successively introduced. Einstein's doubts and thoughts helped me with this.
The principles of Einstein's general theory of relativity were established in 1913 [4]. One of the basic principles - the law of conservation of energy-momentum provided for the conservation of energy, not only the conservation of energy-momentum of matter, but also the energy of the total energy of matter and the gravitational field was not included in the equation of the gravitational field. Einstein showed that choosing with the condition $\sqrt{-g}=1$ simplifies the equation. This brings Einstein's equation (2) to the system of equations [5]. Einstein's equation took on a strange form:
\begin{equation}
\left\{
\begin{array}{c}
\frac{\partial \Gamma_{\mu v}^\alpha}{\partial x^\alpha}-\Gamma_{\mu \beta}^\alpha \Gamma_{v \alpha}^\beta=\kappa\left(T_{\mu v}-\frac{1}{2} g_{\mu \nu} T\right) \\
\sqrt{-g}=1
\end{array}
\right.
\tag{1}
\end{equation}

It was in this form that the equation was solved by Schwarzschild [6]. Due to the complexity of this solution, this variation of Einstein's equation has virtually disappeared from the literature.

At the same time, an equation was obtained that was interpreted as a conservation law. However, it was later found that the components of this equation (pseudotensors) are not components of the energy-momentum of the gravitational field. Despite this, it was possible to obtain energy and momentum for closed systems $[7,8]$.

The strangeness of this equation is that the appearance of a non-covariant condition is not caused by any physical considerations, but only with the sole purpose of "simplifying" the gravitational field equation.

Let us note that the Newtonian limit $\sqrt{-g} \rightarrow 1$, a necessary principle of the general theory of relativity, is also satisfied for the equation without the condition $\sqrt{-g}=1$. If we exclude this condition from equation (1):

\begin{equation}
\frac{\partial \Gamma_{\mu v}^\alpha}{\partial x^\alpha}-\Gamma_{\mu \beta}^\alpha \Gamma_{v \alpha}^\beta=\kappa\left(T_{\mu \nu}-\frac{1}{2} g_{\mu \nu} T\right)
\tag{2}
\end{equation}

This shortened field equation will satisfy all the necessary principles of general relativity if we prove the covariance of the right-hand side of the equation and remember that Einstein had already proven the statement that in the Newtonian limit the equation turns into the Poisson equation.
Thus, equation (2) is a complete equation of the gravitational field, satisfying all the
principles introduced by Einstein into the general theory of relativity.
We will solve equation (2) using numerical methods.

\section{The Schwarzschild problem for a point source}
The Schwarzschild problem for a point source field for a shortened field equation is solved in the same way as the Einstein equation.

We are looking for solutions in the form
\begin{equation}
\mathrm{d} s^2 = \mathrm{s}(\mathrm{r}) \, \mathrm{d} t^2 - \mathrm{p}(\mathrm{r}) \, \mathrm{d} r^2 - r^2 \, \mathrm{d} \theta^2 - r^2 \sin^2 \theta \, \mathrm{d} \varphi^2
\tag{3}
\end{equation}

Solving the shortened equation for empty space
\begin{equation}
\frac{\partial \Gamma_{\mu \nu}^\alpha}{\partial x^\beta} - \Gamma_{\mu \gamma}^\alpha \Gamma_{\nu \beta}^\gamma = 0
\tag{4}
\end{equation}

Nonzero Christoffel symbols of equation (3)
2

$$
\begin{gathered}
\Gamma_{11}^1=\frac{p^{\prime}}{2 p}, \quad \Gamma_{22}^1=\frac{-r}{p}, \quad \Gamma_{33}^1=-\frac{r \sin (\theta)^2}{p} \\
\Gamma_{14}^4=\Gamma_{41}^4=\frac{s^{\prime}}{2 s}, \quad \Gamma_{44}^1=\frac{s^{\prime}}{2 p}, \quad \Gamma_{33}^2=-\sin (\theta) \cos (\theta) \\
\Gamma_{12}^2=\Gamma_{21}^2=\Gamma_{13}^3=\Gamma_{31}^3=\frac{1}{r}, \quad \Gamma_{23}^3=\Gamma_{32}^3=\frac{\cos (\theta)}{\sin (\theta)}
\end{gathered}
$$

From these data and equation (4). Which allows us to find a nonlinear system of ordinary differential equations:
\begin{equation}
\left\{
\begin{array}{r}
\frac{s^{\prime}}{s^{\prime}} - \frac{p^{\prime}}{p} - \frac{s^{\prime}}{s} = 0 \\
\left(\frac{p^{\prime}}{2 p}\right)^{\prime} - \frac{p^{\prime} p^{\prime}}{4 p^2} - \frac{2}{r^2} - \frac{s^{\prime} s^{\prime}}{4 s^2} = 0
\end{array}
\right.
\tag{5}
\end{equation}

where the prime denotes the derivative with respect to r. Eliminating the value $\frac{p^{\prime}}{p}$ from which we obtain a third-order equation:
\begin{equation}
\left(\frac{s^{\prime \prime}}{s^{\prime}} - \frac{s^{\prime}}{s}\right)^{\prime} - \frac{1}{2} \left(\frac{s^{\prime \prime}}{s^{\prime}} - \frac{s^{\prime}}{s}\right)^2 - \frac{4}{r^2} - \frac{s^{\prime} s^{\prime}}{2 s^2} = 0
\tag{6}
\end{equation}

We were unable to find an analytical solution to this equation. A graphical numerical solution is presented in Fig. 1. The resulting solution is a metric in which the parameter $g_{00}$ can be considered as a measure of gravitational potential. In a region remote 1 from the field source, the
solution differs slightly from the Schwarzschild one. when approaching the source, the potential $\left(g_{00}\right)$ quickly decreases to extreme positive values of the order of $10^{-20}$. Moreover, a flat plateau with such values penetrates into the region of the Schwarzschild solution, which has no solutions $r<r_g$ at all. However, the solution by numerical methods is unstable, and the size of the potential well, depending on the position of the boundary conditions, the area of calculations, the algorithm and the number of points, determines the size of the plateau or the location of premature termination of the program. We assume that this is caused by the inadequacy of the Schwarzschild solution as an asyptotic solution for the initial conditions.

When solving the equation numerically, asymptotic proximity to the Schwarzschild solution was used. This made it possible to use the Schwarzschild solution values $1-\frac{1}{r}$ as boundary conditions close to the expected solution. As expected, in the region of relatively small values the solution follows the Schwarzschild solution. Then, similarly to the Schwarzschild solution, it quickly decreases (Fig. 1). But the difference from the Schwarzschild solution, the solution to equation (4) remains positive, although extremely small. A potential well is observed with an almost limiting potential value, which is equal to $\varphi=-c^2$ (Fig. 2).

In Fig. 2 it can be seen that the calculated curve smoothly passes into a plateau. This confirms the rule that proper time must be strictly greater than zero, and singularities in time are unacceptable. The question of the existence of a solution will be helped by solving system (5) for the second component of the metric $g_{11}=p(r)$. To do this, we must exclude $\mathrm{s} \backslash$ left(rivight) from the system of equations (5). From the first equation (5)

$$
\frac{s^{\prime \prime}}{s^{\prime}}-\frac{s^{\prime}}{s}=\frac{p^{\prime}}{p}
$$
This can be integrated with respect to r :

$$
\ln \frac{s^{\prime}}{s}=\ln p+C_1
$$

where $C_1$ is the integration constant. Let's transform this equation

$$
\frac{s^{\prime}}{s}=p e^{c_1}
$$

Let's substitute this result into the second equation of system (5):
\begin{equation}
\left(\frac{p^{\prime}}{2 p}\right)^{\prime} - \frac{p^{\prime} p^{\prime}}{4 p^2} - \frac{2}{r^2} - C_2 \frac{p^2}{4} = 0
\tag{7}
\end{equation}

and $C_2$ is the new constant.
It is clear that it is necessary to know the numerical value of the constant $C_2$ in order to apply numerical methods to solve the equation. This possibility is realized simultaneously with the initial conditions at point $r_0$. Since we assume the asymptotic equality of the solution to equation (7) to the Schwarzschild solution $p_s=r /(r-1)$ the boundary (initial) conditions are determined:

\begin{equation}
\begin{aligned}
r_0 ; \\
p_0 &= p_s\left(r_0\right) ; \\
p_0^{\prime} &= p_s^{\prime}\left(r_0\right) ; \\
p_0^{\prime \prime} &= p_s^{\prime \prime}\left(r_0\right)
\end{aligned}
\tag{8}
\end{equation}

here $r_0$ is the position point of the boundary conditions.
If we substitute boundary conditions (8) into equation (7), we obtain an equation from which we can calculate the value of $C_2$. A simplified value of this quantity can be estimated by eliminating derivatives

$$
C_2=-\frac{8}{p_0^2 r_0^2}
$$

Since in the Newtonian limit $r \rightarrow \infty$, then $C_2=0$. Now we can proceed to the numerical solution of equation (7).
We remember that when solving the Einstein equation, the relation \( g_{11} = (g_{00})^{-1} \) is satisfied. We see something similar here when solving the truncated equation.

\begin{figure}
    \centering
    \includegraphics[width=1\linewidth]{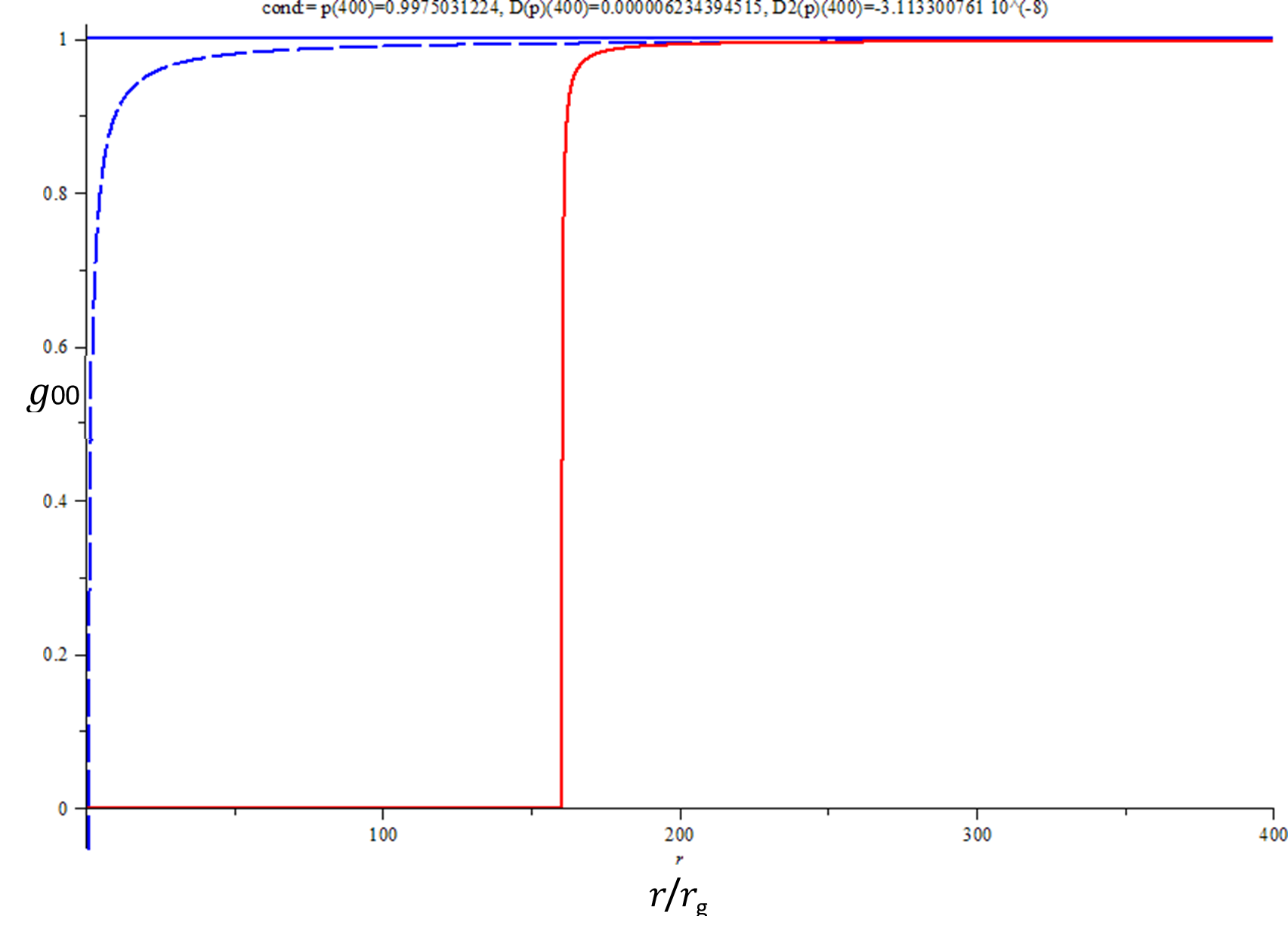}
    \caption{Fig. 1 Dependence of $g_{00}$ on the ratio $\frac{r}{r_g}$. Schwarzschild solutions of the Einstein equation (dotted line) and a numerical solution of the exact gravitational field equation (6) (solid line).}
    \label{fig:enter-label}
\end{figure}
\begin{figure}
    \centering
    \includegraphics[width=1\linewidth]{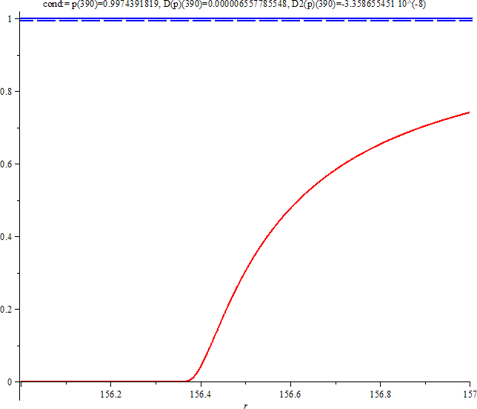}
    \caption{Fig. 2 Enlarged section of Fig. 1, the horizontal section of the solution to equation (8) has a positive value, but so small that the numerical solution program cannot cope and stops.}
    \label{fig:enter-label}
\end{figure}

\begin{figure}
    \centering
    \includegraphics[width=1\linewidth]{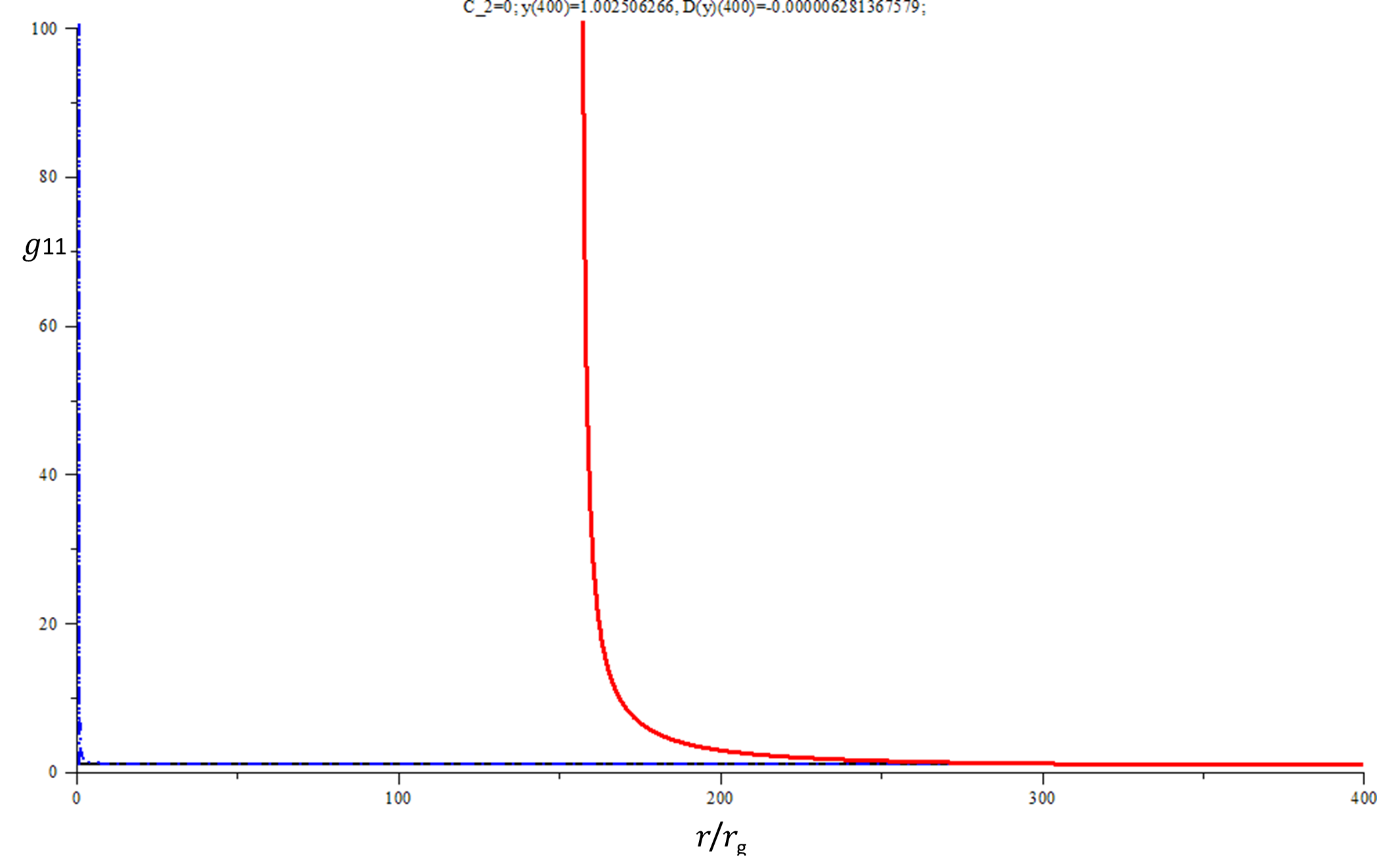}
    \caption{Fig. 3. Dependence of the metric tensor component $g_{11}$ (red line) on the distance to the origin $r / r_g$ in comparison with the Schwarzschild solution (blue dashed line).}
    \label{fig:enter-label}
\end{figure}

\section{Conclusion}
Einstein's empty space has finally received the status of an independent
physical object with parameters specified by the energy-momentum-stress tensor.
The Einstein equation is a good approximation only for not too large gravitational fields, so when calculating extremely large fields it must be replaced by the exact gravitational field equation:

$$
R_{\mu \nu}-\frac{1}{2} g_{\mu \nu} \mathrm{R}=\kappa T_{\mu \nu} \Rightarrow \frac{\partial \Gamma_{\mu \nu}^\alpha}{\partial x_\alpha}-\Gamma_{\mu \beta}^\alpha \Gamma_{v \alpha}^\beta-\frac{1}{2} g_{\mu \nu} \mathrm{R}=\kappa T_{\mu v}
$$
Formally, the inclusion of the tensor \( f_{ab} \) in the Einstein equation leads to a truncated
equation, the solution of which must be a smooth metric tensor. Moreover, the law of conservation
of energy-momentum follows directly from the equation of the gravitational field.
For some applications, solutions with large fields are important. There is confidence that
the fundamental tensor entirely belongs to the original Riemannian space and does not contain
unacceptable values.
In the solutions of the new shortened gravitational field equation, apparently there are no
solutions that go beyond the Riemannian space and that give rise to hypotheses about the structure
of singular objects, like black holes. The structure of the observed compact heavy objects “black
holes” is quite explainable by the presence of potential wells in the solutions in smooth Riemann
space (Fig. 2).
\section*{Declaration of competing interest}
The authors declare that they have no known competing financial interests or personal relationships that could have appeared to influence the work reported in this paper

\section{Attachments}
\section*{Division of the Ricci Tensor}

Following Einstein \cite{einstein1915feldgleichungen}, we represent the Ricci tensor as the sum of two parts:
\[
R_{\mu\nu} = A_{\mu\nu} + B_{\mu\nu},
\]
where
\[
A_{\mu\nu} = \frac{\partial \Gamma_{\mu\nu}^\alpha}{\partial x^\alpha} - \Gamma_{\mu\beta}^\alpha \Gamma_{\nu\alpha}^\beta,
\]
\[
B_{\mu\nu} = -\frac{\partial \Gamma_{\mu\alpha}^\alpha}{\partial x^\nu} + \Gamma_{\mu\nu}^\alpha \Gamma_{\alpha\beta}^\beta.
\]

However, we do not yet know whether parts of the Ricci tensor are tensor quantities. The following theorem allows us to prove this.

\section*{Theorem 1}

The magnitude of the simplified Christoffel symbol $\Gamma_{\mu\alpha}^\alpha$ is a 4-vector. $A_{\mu\nu}$ and $B_{\mu\nu}$ are tensors.

\paragraph{Proof.}
We use the formula for the coordinate transformation of Christoffel symbols \cite{Reference8}:
\[
\Gamma_{\mu\alpha}^\iota = \Gamma_{\nu\xi}'^\gamma \frac{\partial x^\iota}{\partial x'^\gamma} \frac{\partial x'^\nu}{\partial x^\mu} \frac{\partial x'^\xi}{\partial x^\alpha} + \frac{\partial^2 x'^\gamma}{\partial x^\alpha \partial x^\mu} \frac{\partial x^\iota}{\partial x'^\gamma}.
\]
Setting $\iota = \alpha$ and $\gamma = \xi$, the formula simplifies:
\[
\Gamma_{\mu\alpha}^\alpha = \Gamma_{\nu\xi}'^\xi \frac{\partial x'^\nu}{\partial x^\mu} + \frac{\partial^2 x'^\xi}{\partial x^\mu \partial x^\alpha} \frac{\partial x^\alpha}{\partial x'^\xi}.
\]

The last term simplifies using the theorem on the equality of mixed derivatives. It turns out to be zero, leading to:
\[
\Gamma_{\mu\alpha}^\alpha = \Gamma_{\nu\xi}'^\xi \frac{\partial x'^\nu}{\partial x^\mu}.
\]

Thus, $\Gamma_{\mu\alpha}^\alpha$ transforms as a 4-vector. Covariant differentiation generates the tensor:
\[
\frac{D\Gamma_{\mu\alpha}^\alpha}{\partial x^\nu} = \frac{\partial \Gamma_{\mu\alpha}^\alpha}{\partial x^\nu} - \Gamma_{\mu\nu}^\alpha \Gamma_{\alpha\beta}^\beta,
\]
which corresponds to $B_{\mu\nu}$. Since $R_{\mu\nu}$ is a tensor and $R_{\mu\nu} = A_{\mu\nu} + B_{\mu\nu}$, it follows that $A_{\mu\nu}$ is also a tensor.

\section*{Connection of $B_{\mu\nu}$ with the Gravitational Field}

In the Newtonian approximation, described by the stationary metric:
\[
ds^2 = (1 + 2\phi/c^2) dt^2 - dx^2 - dy^2 - dz^2,
\]
where $\phi(x)$ is the Newtonian potential, the spatial components of the metric tensor slightly differ from unity. We obtain:
\[
B_0^0 = \Gamma_{00}^1 \Gamma_{10}^0 = \frac{1}{c^2} \frac{\partial \phi}{\partial x} \cdot \frac{1}{c^2 + 2\phi} \frac{\partial \phi}{\partial x} \approx \frac{1}{c^4} \left( \frac{\partial \phi}{\partial x} \right)^2.
\]

This differs in sign from the known field density in the Newtonian approximation \cite{Reference8}:
\[
f = -\frac{(\nabla \phi)^2}{8\pi G}.
\]

By reversing the energy scale, the energy density becomes positive:
\[
B_{00} = \frac{8\pi G}{c^4} f_{00} = -\kappa f_{00} = \kappa f'_{00}.
\]

This establishes the relation:
\[
B_{\mu\nu} = \kappa f'_{\mu\nu}.
\]

\section*{Theorem 2}

As the determinant of the metric tensor $g$ approaches the value of the Minkowski metric, equation (1) approaches Einstein's equation (2).

\paragraph{Proof.}
As shown, $B_{\mu\nu} \to 0$ as $g_{\mu\nu} \to \eta_{\mu\nu}$, so:
\[
\frac{\partial \Gamma_{\mu\nu}^\alpha}{\partial x^\alpha} - \Gamma_{\mu\beta}^\alpha \Gamma_{\nu\alpha}^\beta \approx \frac{\partial \Gamma_{\mu\nu}^\alpha}{\partial x^\alpha} - \Gamma_{\mu\beta}^\alpha \Gamma_{\nu\alpha}^\beta + B_{\mu\nu} = R_{\mu\nu}.
\]

This establishes the asymptotic equivalence of equation (1) and Einstein's equation.

\section*{Theorem 3}

From the gravitational field equation (1), the complete law of conservation of matter and the gravitational field follows.

\paragraph{Proof.}
The covariant derivative of the Einstein tensor is:
\[
G^\mu_{\nu;\mu} = 0.
\]

Hence, from equation (2), the general conservation law is:
\[
T^\mu_{\nu;\mu} + f^\mu_{\nu;\mu} = 0.
\]

\end{document}